\documentclass[aps,pra,superscriptaddress,amssymb,showpacs,twocolumn]{revtex4-1}
\usepackage{amsfonts} 
\usepackage{amsmath}
\usepackage{amssymb}
\usepackage{graphicx}
\usepackage{hyperref}
\usepackage{color}
\usepackage{natbib}  
\usepackage{bm} 
\usepackage{isomath} 
\usepackage{braket}

\DeclareSymbolFont{bbgreek}{U}{bbold}{m}{n}

\providecommand{\U}[1]{\protect\rule{.1in}{.1in}}

\newcommand{\be}{\begin{equation}}
\newcommand{\ee}{\end{equation}}

\newcommand{\vect}[1]{\vectorsym{#1}} % vectors
\begin{document}

\title{Hierarchies of multipartite entanglement for continuous-variable states}

\author{Antonio A. Valido}
\email{aavalido@ull.es}
\affiliation{Instituto Universitario de Estudios Avanzados (IUdEA) and Departamento de F\'{\i}sica, Universidad de La Laguna, La Laguna 38203 Spain}
\author{Federico Levi}
\affiliation{Freiburg Institute for Advanced Studies, Albert-Ludwigs University of Freiburg, Albertstra\ss e 19, 79104 Freiburg, Germany}
\author{Florian Mintert}
\affiliation{Freiburg Institute for Advanced Studies, Albert-Ludwigs University of Freiburg, Albertstra\ss e 19, 79104 Freiburg, Germany}
\affiliation{Department of Physics, Imperial College London, London SW7 2AZ, United Kingdom}


\begin{abstract}
We derive a hierarchy of separability criteria for multi-mode continuous variable systems. They permit to study in a unified way the $k$-partite entanglement of broad classes of Gaussian and non-Gaussian states. With specific examples we demonstrate the strength of the criteria, and, we discuss their assessment based on data obtained from Gaussian measurements.

\end{abstract}

\date{\today}

\maketitle
\section{Introduction}

Entanglement has proven to be a central resource in quantum information processing using either discrete or continuous variable (CV) systems (such as field modes of light, nanomechanical oscillators or cold atomic gases)  \cite{horodecki20091}.
Any attempt to create an entangled state is limited by the residual noise and decoherence,
and proper tools to verify entanglement are needed to evidence the success of an experiment.
In CV systems these tools can roughly be divided into those that apply to Gaussian states
\cite{duan20001,simon20001,werner20001,giedke20011,adesso20061} (see \cite{adesso20071} for a complete review),
and those that apply to more general states \cite{gabriel20111,jiang20141}.
Most tools entail an optimization of an entropy-like functional like a convex roof construction \cite{wolf20041,adesso20121},
the proper choice of a set of observables that witness the entanglement for a broad class of states \cite{hyllus20061,walborn20091,saboia20111,nha20121,zhang20131,sperling20131},
or the suitable selection of a finite \cite{vanloock20031,agarwal20051,serafini20061,hillery20061,nha20061,sun20091} or infinite \cite{shchukin20051,nha20081} series of inequalities (concerning moments of the quadrature variables) which are mainly based on the well-known criterion of positive partial transposition (PPT) \cite{horodecki20091,eisert20031}. The need to optimize or accurately choose a tool in accordance with the specific properties of a quantum state makes the characterization of entanglement a computationally intricate problem \cite{huang20141}, which becomes even more involved as the mixedness of the state or the number of constituents of the system grows.

Entanglement shared by two subsystems has been realized experimentally in various systems \cite{li20131},
but increasing the number of entangled components is a big experimental challenge, such that the preparation of states with more than bipartite entanglement has been achieved in few systems only \cite{leibfried20051,haffner20051,shalm20131}.
The limitations due to noise and decoherence typically get increasingly severe with growing number of entangled subsystems.
Under given imperfect conditions it might not be possible to create a genuinely $n$-partite entangled state in an $n$-partite system,
whereas the preparation of a bipartite entangled state might still be feasible.
Tools to verify bipartite or genuine $n$-partite entanglement have been explored in detail \cite{bourennane20041},
but tools that analyse the range in between have been established only recently \cite{guhne20051,gabriel20111,vitagliano20111,levi20131,huber20131}.
Only those tools, however, will help us to gauge experimental progress and eventually achieve the creation of genuine $n$-partite entanglement.

We build up here on a hierarchy of separability criteria that detect $k$-partite entanglement in $n$-partite discrete systems \cite{levi20131},
and extend this approach to the case of contiuous variable systems. Based on this hierarchy, we present versatile hierarchies of separability criteria that apply to Gaussian and non-Gaussian states such as photon-added/subtracted states \cite{kim20081} that  display particularly strong non-classical correlations properties \cite{kitagawa20061,yang20091,navarrete20121,bartley20131,kim20131,chowdhury20141}.

The paper is organized as follows: We start with an introduction to CV systems and hierarchies of separability criteria in Sec. \ref{SHIC}. The formulation of these hierarchies for CV systems is presented in Sec. \ref{SHCV}, which is accompanied with a discussion of the similarities with the PPT criterion (see Sec. \ref{SRPPT}). We apply these hierarchies to Gaussian and non-Gaussian states in Sec. \ref{SExa}, and the possible experimental assessment of the criterion is discussed in Sec. \ref{SEAS}.  

%%%%%%%%%%%%%%%%%%%%%%%%%%%%%%%%%%%%%%%%%%%%%%%%%%%%%%%%%%%%%%%%%%%
\section{Basic definitions}\label{SHIC}

%%%%%%%%%%%%%%%%%%%%%%%%%%%%%%%%%%%%%%%%%%%%%%%%%%%%%%%%%%%%%%%%%%%
\subsection{Phase space representation}
The Hilbert space $\mathcal{H}_{n} $ of a quantum system composed by $n$ modes results from the $n$-fold tensor product of the single-mode Hilbert space $\mathcal{H}_{1}=\mbox{L}^{2}(\mathbb{R})$, and all the physical information
about the system is encoded in the density operator $\hat{\rho}$. The $m$-th mode is described in terms of the canonical operators,
{\it i.e.} position $\hat{Q}_{m}$ and momentum $\hat{P}_{m}$.
Equivalently it may be described by their dimensionless counterparts $\hat{q}_{m}=\hat{Q}_{m}\sqrt{M\Omega/\hbar }$ and $\hat{p}_{m}=\hat{P}_{m}/\sqrt{M\Omega\hbar} $ defined in terms of the frequency $\Omega$ and and mass $M$.
From now on we will use only the dimensionless operators
and define the operator-valued vector $\hat{\vect x}=(\hat{q}_{1},\hat{p}_{1},...,\hat{q}_{n},\hat{p}_{n})^{T}$ whose elements satisfy the canonical commutation relations $\left[\hat{\vect x}_{m},\hat{\vect x}_{l} \right]=-i \left[ \vect J_{n}\right] _{ml}  $, with the symplectic matrices
\begin{equation}
\vect J_{n}=\bigoplus_{m=1}^{n}\vect J_{1}\quad \text{and} \quad \vect J_{1}=\left(\begin{array}{cc}
0 & -1 \\
1 &  0 \\
\end{array} \right),
\nonumber
\end{equation}
of the composite system and a single subsystem.

It is convenient to describe a continuous variable system in terms of the real symplectic space $(\mathbb{R}^{2n},\vect J_{n})$, {\it i.e.} phase space \cite{eisert20031,adesso20071},
rather than the infinite dimensional complex Hilbert space $\mathcal{H}_{n}$.
Quantum mechanical operators $\hat{A}$ are then replaced by their Weyl symbol
\begin{eqnarray}
W_A(\vect x)&=&\int_{\mathbb{R}^{2n}}\frac{d^{2n}\vect \xi}{(2\pi)^{2n}}e^{i\vect x^{T}\vect J_{n} \vect \xi}Tr\left[\hat{A} e^{-i  \hat{\vect x}^{T} \vect J_{n} \vect \xi} \right],
\label{Wsymbol}
\end{eqnarray}
{\it i.e.} functions $W_A(\vect x)$ of classical phase space variables $\vect x=(q_{1},p_{1},...,q_{n},p_{n})$ \cite{perelomov1986}.
The Weyl symbol of a density matrix $\hat\rho$ is typically referred to as Wigner function, and it is denoted by 
$W(\vect x)$ \cite{weedbrock20121}.

The Wigner function $W (\vect x)$ of a Gaussian state $\hat{\rho}$ has the particularly simple form \cite{weedbrock20121}
\begin{equation}
W (\vect x)=\frac{e^{-\frac{1}{2}(\vect x-\bar{\vect x})^{T} \vect V^{-1}(\vect x-\bar{\vect x})}}{(2\pi)^{n}\sqrt{\det(\vect V ) }},
\nonumber 
\end{equation}
where the vector $\bar{\vect x}=\mbox{Tr}(\hat{\rho}\hat{\vect x}) $ contains the expectation values (first-moments) of the dimensionless phase space variables,
and the covariance matrix $\vect V$ is defined by
\begin{equation}\nonumber
\vect V_{ml}=\frac{1}{2}\mbox{Tr}\left( \hat{\rho}\left\lbrace \left[ \hat{\vect x}\right] _{m}-\left[ \bar{\vect x}\right] _{m},\left[ \hat{\vect x}\right]_{l}-\left[ \bar{\vect x}\right] _{l}\right\rbrace \right)\ ,
\end{equation}
where $\lbrace .,.\rbrace$ denotes the anti-commutator. In this case, $W$ is completely characterized by the vector $\bar{\vect x}$ and the real symmetric $2n\times 2n$ matrix $\vect V$, {\it i.e.} by $2n^{2}+n$ real parameters.
According to the Heisenberg uncertainty relation, the covariance matrix of any quantum state must satisfy $ \vect V \geq \frac{i}{2} \vect J_{n}$ \cite{eisert20031,weedbrock20121}, which implies the positive definiteness $\vect V > 0$.
Since the entanglement of the system is invariant under  local unitary displacements \cite{eisert20031}, we shall take the first-moment vector equal to zero ($\bar{\vect x}=0$) from now on. 

Here, we are concerned with the class of entangled states $\hat{\rho}$ whose Wigner function may be expressed as the product of a polynomial function $F(\vect x)$ and the Wigner function of a Gaussian state with covariance matrix $\vect V$, {\it i.e.} 
\begin{equation}
W(\vect x)=\frac{F(\vect x)e^{-\frac{1}{2}\vect x\vect V^{-1} \vect x}}{(2\pi)^{n}\sqrt{\det(\vect V ) }}.
\label{WignerfG}
\end{equation}
Direct examples of this kind of states are those states which are generated by a series of photon-creation \cite{agarwal20111} or photon-subtraction operations \cite{hu20121,xu20131,chowdhury20141}, or more general, a coherent superposition of both \cite{agarwal20051,guo20131}. We shall refer to the latter as photon-manipulated states. In that case, the degree of the polynomial corresponds to the number of such manipulations that need to be applied to a Gaussian state to arrive at the state in question.
We should, however, stress that $F(\vect x)$ may be also an analytic function with domain in all the phase space (a function with a convergent Taylor series), such that the set of non-Gaussian states with Wigner function (\ref{WignerfG}) may comprise a broader class of CV states than the photon-manipulated states, as for example Schr\"{o}dinger cat states.

%%%%%%%%%%%%%%%%%%%%%%%%%%%%%%%%%%%%%%%%%%%%%%%%%%%%%%%%%%%%%%%%%%%%%%%%%%%%%%%%
\subsection{Hierarchy of separability criteria} \label{SH}

A pure state of an $n$-partite quantum system is considered $n$-partite entangled
if it can not be written as a simple tensor product of two state-vectors each of which describes a part of the subsystems only.
If an $n$-partite quantum state can not be written as a simple tensor product of $k_i$-partite entangled $k_i$-partite state-vectors with $k_i<k$,
then the state is $k$-partite entangled.

A mixed $n$-partite state $\hat\rho$ is considered $k$-partite entangled if it can not be represented as an average over projectors onto pure states that are less than $k$-partite entangled, {\it i.e.}
\begin{equation}
\hat{\rho}\neq \sum_{j=1}^{k-1}\int d\mu_j(a)\left|\Psi_{j,n}^{(a)}\right\rangle \left\langle \Psi_{j,n}^{(a)}\right|,
\label{Ssep}
\end{equation}
where $ \ket{\Psi_{j,n}^{(a)}}$ are $j$-partite entangled $n$-partite states, $\mu_j(a)$ are positive functions that satisfy $\sum_{j=1}^{k-1}\int d\mu_j(a)=1$, and the summation is restricted to values $j< k$. Physically, this definition means that a $k$-partite entangled state can be realized by mixing different states that are at most $k$-partite entangled, but since the states that enter this average may carry entanglement between different groups of subsystems, a $k$-partite entangled $n$-partite state is not necessarily separable with respect to a certain bipartition.

Our starting point to detect $k$-partite entanglement is a hierarchy of separability criteria $\tau_{k,n}$. It is based on a comparison between several matrix elements of the density operator in question with respect to some product states. As shown in \cite{huber20101}, genuine $n$-partite entanglement is identified through the condition
\begin{eqnarray}
\tau_{n}(\hat{\rho})&=&\underbrace{\left|\bra{\Phi_1}\varrho\ket{\Phi_2}\right|}_{f(\varrho)} \nonumber \\
&-&\sum_{j=1}^{2^{n-1}-1}\underbrace{\sqrt{\bra{\Phi_{1j}}\varrho\ket{\Phi_{1j}}\bra{\Phi_{2j}}\varrho\ket{\Phi_{2j}}}}_{f_j(\varrho)}>0\ ,
\label{dummy1}
\end{eqnarray}
where $\ket{\Phi_1}=\bigotimes_{m=1}^{n}\ket{\varphi_m}$ and $\ket{\Phi_2}=\bigotimes_{m=1}^{n}\ket{\varphi_{n+m}}$ are two product vectors,
and the vectors $\ket{\Phi_{1i}}$ and $\ket{\Phi_{2i}}$ are defined in terms of the inequivalent possibilities to divide the $n$-subsystems into two groups: there are $2^{n-1}-1$ inequivalent such bipartitions, each of which that can be characterized by a vector $\vect v_j$ whose $n$ elements adopt the values $0$ or $1$, and the groups are defined by the subsystems associated with the value $0$ and $1$ respectively. In terms of these vectors, we have the definition
\be
\ket{\Phi_{1j}}=\bigotimes_{m=1}^{n}\ket{\varphi_{m+n \left[ \vect v_{j}\right] _{m}}}\ ,\ket{\Phi_{2j}}=\bigotimes_{m=1}^{n}\ket{\varphi_{m+n-n\left[ \vect v_{j}\right] _{m}}}\ ,
\label{dummy2}
\ee
that is, the vectors $\ket{\Phi_{1j}}$ and $\ket{\Phi_{2j}}$ are obtained from the vectors  $\ket{\Phi_{1}}$ and $\ket{\Phi_{2}}$ through a permutation of state vectors $\ket{\varphi_{m}}$ with $\ket{\varphi_{n+m}}$ that belong to those subsystems that are grouped together in the $j$-th bipartition.

If a pure state $\hat{\rho}=\left|\Psi\right\rangle \left\langle \Psi\right|$ is separable with respect to the $j$-th bipartition, then $f(\hat{\rho})=f_j(\hat{\rho})$.
Since the $f_j(\hat\varrho)$ are non-negative, this implies that $\tau_n$ is non-positive.
As this reasoning holds for any bipartition, and, in addition $\tau_n$ is convex,
$\tau_n$ is indeed non-positive for any state $\varrho$ that can be decomposed into bi-separable pure states. 

A fully separable pure state is bi-separable with respect to all bi-partitions; accordingly, one may introduce the function
$\tau_{bi,n}(\hat{\rho})=f(\varrho)-(2^{n-1}-1)^{-1}\sum_{j=1}^{2^{n-1}-1}f_j(\varrho)$,
and a positive value of $\tau_{bi,n}$ identifies a mixed state to be at least bi-partite entangled. In the same fashion, one can introduce scalar factors $a_{j}^{(k,n)} \geq 0$ \cite{levi20131} for $n\geq k\geq2$ such that
\begin{equation}
\tau_{k,n}(\hat{\rho})=f(\varrho)-
\sum_{j}a_{j}^{(k,n)} f_{j}(\varrho)
\label{Tau}
\end{equation}
can be positive only if $\hat{\rho}$ is at least $k$-partite entangled.

In order to detect entanglement properties as reliably as possible, a suitable choice of {\em probe vectors} $\ket{\varphi_i}$ is in order.
In practice, it is desirable to find an optimal set of normalized such vectors that maximize $\tau_{k,n}$.
Advantageously, the number of probe vectors scales only linearly with $n$,
but a full optimization over the infinite-dimensional vectors without simplifying assumptions does not seem to be a fruitful endeavour.
Similarly to the concept of Gaussian entanglement of formation \cite{wolf20041},
we therefore require that all probe vectors are Gaussian.
Each Gaussian probe state $ \ket{\varphi_{m}}$ is then characterized by it first and second moments
\begin{eqnarray}
\bar{\vect x}_{m}&=&(\bar{q}_{m},\bar{p}_{m}), \label{Xvector} \mbox{ and}\\
\vect \Sigma_{m} &=&
\left[ \begin{array}{cc}
     \sigma_{xx}^{(m)} & \sigma_{xp}^{(m)}\\
      \sigma_{xp}^{(m)}  & \sigma_{pp}^{(m)} \\
\end{array}\right],\label{MCV}
\end{eqnarray}
with $\det(\vect \Sigma_{m})=1/4$, $\sigma_{xx}^{(m)}\ge 0$ and $\sigma_{pp}^{(m)}\ge 0$.
In the following we will identify choices for these parameters that yield strong criteria.
Remarkably enough, this allows us to reproduce the PPT criterion for two-mode and pure three-mode Gaussian states.
Beyond that, even with this simplifying assumption, Eq.\eqref{Tau}, is able to detect non-Gaussian entanglement \cite{agarwal20051}, for which criteria only based on the second moments of the quadrature variables fail.
Both observations demonstrate that assuming Gaussian probe states, makes the present hierarchy an easily accessible but strong tool.

%%%%%%%%%%%%%%%%%%%%%%%%%%%%%%%%%%%%%%%%%%%%%%%%%%%%%%%%%%%%%%%%%%%
\section{Hierarchies of Inseparability Criteria for CV systems}\label{SHCV}

The $\tau_{k,n}$ are parametrized by the first and second moments of the Weyl symbols of the operators $\Ket{\Phi_{1}}\Bra{\Phi_{1}} $, $\Ket{\Phi_{2}}\Bra{\Phi_{2}} $, $\Ket{\Phi_{1j}}\Bra{\Phi_{1j}} $ , and $\Ket{\Phi_{2j}}\Bra{\Phi_{2j}} $.
Let us denote their vectors of first moments by $\vect X_{\Phi_{1}}$, $\vect X_{\Phi_{2}}$, $\vect X_{\Phi_{1j}}$ and $\vect X_{\Phi_{2j}}$,
and their matrices of second moments by
$\vect \Sigma_{\Phi_{1}}$, $\vect \Sigma_{\Phi_{2}}$, $\vect \Sigma_{\Phi_{1j}}$ and $\vect \Sigma_{\Phi_{2j}}$.
Since also the matrix element $\bra{\Phi_1}\hat\rho\ket{\Phi_2}$ enters the definition of $\tau_{k,n}$,
it is convenient to introduce also moments
\begin{equation}
\vect X_{\Phi_{21}}=\frac{\int d^{2n}\vect x \  \vect x\ W_{\ket{\Phi_2}\bra{\Phi_1}}(\vect x)}{\int d^{2n}\vect x\ W_{\ket{\Phi_2}\bra{\Phi_1}}(\vect x)}
\label{Def}
\end{equation}
and $\vect \Sigma_{\Phi_{12}}$ defined analogously,
where the explicit normalization is introduced because the overlap between $\ket{\Phi_1}$ and $\ket{\Phi_2}$ is typically not unity.

As shown in Eq.(\ref{AWsymbol}) in the appendix \ref{ATool}, $\vect \Sigma_{\Phi_{21}}$ can easily be constructed from the covariance matrices $\vect\Sigma_{m}$ defined in Eq.~\eqref{MCV} via the prescription
\begin{eqnarray}
\vect \Sigma_{\Phi_{21}}&=&\bigoplus^{n}_{m=1}\vect \Sigma_{m,n+m},   \label{CV3}
\end{eqnarray}
with
\begin{eqnarray}
\vect \Sigma_{m,n+m}&=&\frac{\vect \Sigma_{m}+\vect \Sigma_{n+m}}{2\det(\vect \Sigma_{m}+\vect \Sigma_{n+m})}  \nonumber \\
&+&i\frac{\vect \Sigma_{m}\vect J_{1}^{T} \vect \Sigma_{n+m}-\vect \Sigma_{n+m}\vect J_{1}^{T} \vect \Sigma_{m}}{2\det(\vect \Sigma_{m}+\vect \Sigma_{n+m})}\ .
\nonumber
\end{eqnarray}
The first moments are then given by \cite{mintert20091}
\begin{equation}
\vect X_{\Phi_{21}}=\frac{\vect X_{\Phi_{1}}+\vect X_{\Phi_{2}}}{2}+i\vect \Sigma_{\Phi_{21}}\vect J_{n}(\vect X_{\Phi_{1}}-\vect X_{\Phi_{2}})\ .
\label{X1} 
\end{equation}

As it is extensively illustrated in appendix \ref{ATool}, one may express $\tau_{k,n}$ in a rather compact form

\begin{eqnarray}
\tau_{k,n}(\hat{\rho})&=&\frac{e^{-\frac{\alpha}{2} }\left|\mathsf{f}_{\Phi_{21}}\right| }{\sqrt[4]{\det\left(\vect\Sigma_{\Phi_{1}}+ \vect\Sigma_{\Phi_{2}}\right) }}-\sum_{j}a^{(k,n)}_{j}e^{-\frac{\beta_{j}}{4} }\sqrt{\mathsf{f}_{\Phi_{1j}} \mathsf{f}_{\Phi_{2j}}}\ ,
\label{wholetauNG}
\end{eqnarray}
with
\begin{equation}
\mathsf{f}_{u}=\frac{ \exp\left( \frac{1}{2}
K^T
\left( \vect V^{-1}+\vect \Sigma_{u}^{-1}\right)^{-1}
K \right)  F(\vect x) \Big |_{\vect x=\vec{0}}}{\sqrt{\det\left(\vect \Sigma_{u}+\vect V \right) }},
\label{TauNG1}
\end{equation}

and $K=\left( \frac{\partial}{\partial \vect x}+\vect \Sigma_{u}^{-1} \vect X_{u}\right)$
for $u=\Phi_{21}, \Phi_{1j}, \Phi_{2j}$. The quantities
\begin{eqnarray}
\alpha &=&\Re\left(\vect X_{\Phi_{21}}^{T} \vect \Sigma_{\Phi_{21}}^{-1}  \vect X_{\Phi_{21}} \right)  \nonumber          \\
&+&(\vect X_{\Phi_{1}}-\vect X_{\Phi_{2}})^{T}  \vect J_{n}^{T}\Re(\vect \Sigma_{\Phi_{21}} )\vect J_{n} (\vect X_{\Phi_{1}}-\vect X_{\Phi_{2}})
\label{alpha}
\end{eqnarray}
and
\begin{equation}
\beta_{j} = \vect X_{\Phi_{1j}}^{T} \vect \Sigma_{\Phi_{1j}}^{-1}  \vect X_{\Phi_{1j}}+  \vect X_{\Phi_{2j}}^{T} \vect \Sigma_{\Phi_{2j}}^{-1}  \vect X_{\Phi_{2j}}   \ ,
\label{beta}
\end{equation}
are quadratic functions of the first-moment vectors, and $\Re$ denotes the real part. We provide the expressions for the vectors $\vect X_{\Phi_{1}}$, $\vect X_{\Phi_{2}}$, $\vect X_{\Phi_{1j}}$, and $\vect X_{\Phi_{2j}}$, as well as for the covariance matrices $\vect \Sigma_{\Phi_{1}}$, $\vect \Sigma_{\Phi_{2}}$, $\vect \Sigma_{\Phi_{1j}}$, and $\vect \Sigma_{\Phi_{2j}}$ in Eqs. (\ref{Xm1}) to (\ref{X3}), in appendix \ref{APerm}.

The general expression Eq.(\ref{wholetauNG}) holds for any state whose Wigner function can be cast in the form of Eq.(\ref{WignerfG}). If $F(\vect x)=1$ in Eq.(\ref{WignerfG}), {\it i.e.} if $\hat\rho$ is Gaussian, then $\mathsf{f}_{u}$ defined in Eq.~\eqref{TauNG1} takes the simpler form
\begin{equation}
\mathsf{f}_{u}^{(G)}=\frac{ \exp\left( \frac{1}{2}\left(\vect X_{u}\right)^{T} \vect \Sigma_{u}^{-1} \left( \vect V^{-1}+\vect \Sigma_{u}^{-1}\right)^{-1} \vect \Sigma_{u}^{-1} \vect X_{u} \right) }{\sqrt{\det\left(\vect \Sigma_{u}+\vect V \right) }}.
\nonumber
\end{equation}

In order to identify general properties of the states $\ket{\Phi_i}$ that yield potentially maximal values for $\tau_{k,n}$, we will make the assumption

\begin{eqnarray}
\vect \Sigma_{\Phi_{1}}&=&\vect \Sigma_{\Phi_{2}}= \vect \Sigma,
\label{Ass1}
\end{eqnarray}
{\it i.e.} we assume that
$\ket{\varphi_{m}}$ and $\ket{\varphi_{n+m}} $ (for $m=1,...,n$) have the same covariance matrix. 
With this assumption Eqs. (\ref{alpha}) and (\ref{beta}) reduce to $\alpha=\alpha'$ and $\beta_{j}=\beta'_{j}$ with
\begin{equation}
\beta'_{j}=2\alpha'+\frac{1}{2}(\vect X_{\Phi_{1}}-\vect X_{\Phi_{2}})^{T} \vect P_{j}^{T}\vect \Sigma^{-1}\vect P_{j}(\vect X_{\Phi_{1}}-\vect X_{\Phi_{2}})\ ,
\nonumber
\end{equation}
and $\alpha'=1/4(\vect X_{\Phi_{1}}+\vect X_{\Phi_{2}})^{T}\vect \Sigma^{-1}(\vect X_{\Phi_{1}}+\vect X_{\Phi_{2}}) $, with 
\begin{equation}
\vect P_{j}=\bigoplus_{m=1}^{n}(-1)^{\left[ \vect v_{j}\right] _{m}}\vect I\ ,
\label{Pmatrix}
\end{equation}
where $\vect I$ is the two-dimentional identity matrix, and $\vect v_j$, which is defined in the context of Eq.~\eqref{dummy2}, characterizes the bipartition $j$. With the help of the following identity valid for quadratic matrices \cite{matrixI}
\begin{equation}
\frac{1}{\vect \Sigma + \vect V}=\vect \Sigma^{-1} -\vect \Sigma^{-1}(\vect V^{-1}+\vect \Sigma^{-1})\vect \Sigma^{-1},
\end{equation}
one may easily show that the hierarchy $\tau'_{k,n}$ resulting from the assumption Eq.(\ref{Ass1}) can be expressed as
\begin{equation}
\tau_{k,n}'(\hat{\rho})=\frac{e^{-\frac{1}{8}(\vect X_{\Phi_{1}}+\vect X_{\Phi_{2}})^{T}\frac{1}{\vect \Sigma+\vect V}(\vect X_{\Phi_{1}}+\vect X_{\Phi_{2}})}}{\sqrt{\det\left(\vect \Sigma+\vect V \right) }}h_{k,n},
\nonumber
\end{equation}
where $h_{k,n}$ is a function which does not depend on $(\vect X_{\Phi_{1}}+\vect X_{\Phi_{2}})$, {\it i.e.} $h_{k,n}=h_{k,n}(\vect \Sigma,\vect X_{\Phi_{1}}-\vect X_{\Phi_{2}})$.

Since $\vect \Sigma$ and $\vect V$ are positive definite, the exponent is non-positive, such that $\tau_{k,n}'$ adopts its maximum only if $\vect X_{\Phi_{1}}+\vect X_{\Phi_{2}}=0$.
That is, assuming Gaussian probe vectors and  Eq.~\eqref{Ass1} permits to perform an essential part of the maximization of $\tau_{k,n}$ analytically, which eases the reliable estimation of ${\cal T}_{k,n}= \smash{\displaystyle\max_{\Phi_{1},\Phi_{2}}} \tau_{k,n}$ substantially. With this, we arrive at
\begin{equation}
\tilde{{\cal T}}_{k,n}= \displaystyle\max_{\vect X, \vect \Sigma} \tilde{\tau}_{k,n},
\nonumber
\end{equation}
with
\begin{eqnarray}
\tilde{\tau}_{k,n}(\hat{\rho})&=&  \frac{e^{-2\vect X^{T}\vect J_{n}^{T} \frac{1}{\vect \Sigma^{-1} +\vect V^{-1}} \vect J_{n}\vect X }}{\sqrt{\det\left(\vect \Sigma +\vect V \right) }} \nonumber \\
&-&\sum_{j}a^{(k,n)}_{j}\frac{e^{-\frac{1}{2}\vect X^{T}(\vect P_{j})^{T} \frac{1}{\vect \Sigma +\vect V}\vect P_{j}\vect X }}{\sqrt{\det\left(\vect \Sigma +\vect V \right) }}\ ,
\label{Stau}
\end{eqnarray}
which can readily be optimized numerically.

\subsection{Resemblance to the PPT Criterion}\label{SRPPT}

Since Eq.~\eqref{Stau} is the result of several restrictions that potentially weaken the hierarchy, a critical assessment of its strength is in order.
Since most of existing separability criteria are concerned with separability with respect to a given bipartition, we focus for the moment on this question.
According to Eq.~\eqref{Stau}, the inequality
\begin{equation}
e^{-\frac{1}{2}\vect X^{T}(\vect P_{j})^{T} \frac{1}{\vect \Sigma +\vect V}\vect P_{j}\vect X }\geq e^{-2\vect X^{T}\vect J_{n}^{T} \frac{1}{\vect \Sigma^{-1} +\vect V^{-1}} \vect J_{n}\vect X },
\nonumber
\end{equation}
is satisfied for any mixed Gaussian state that is biseparable with respect the bipartition $j$.
Since this scalar inequality is satisfied for any choice of $\vect X$,
it implies the matrix inequality
\cite{horn1985},
\begin{equation}
4\vect J_{n}^{T}\frac{1}{\vect \Sigma^{-1} +\vect V^{-1}} \vect J_{n}\geq (\vect P_{j})^{T}\frac{1}{\vect \Sigma +\vect V} \vect P_{j}\ .
\label{pptR}
\end{equation}
In the following, we will show that this permits us to recover the ppt-criterion for mixed two-mode and pure three-mode Gaussian states, when all the probe states $\ket{\varphi_{m}}$ are chosen to be pure infinitely-squeezed states, with covariance matrix with $\sigma_{pp}^{m} \to 0$ ($\forall m$) for squeezing in momentum, or $\sigma_{xx}^{m} \to 0$ ($\forall m$) for squeezing in position. It is worthwhile noting that if inequality (\ref{pptR}) is violated in all the bipartitions, then $\hat{\rho}$ is genuine multipartite entangled.

%%%%%%%%%%%%%%%%%%%%%%%%%%%%%%%%%%%%%%%%%%%%%%%%%%%%%%%%%%%%%%%%%%
\subsubsection{Two-mode Case}

The covariance matrix $\vect V$ of any two-mode Gaussian state can be expressed in the standard form (\ref{APPTSF1}), in terms of four coefficients $a,b,c,d \in \mathbb{R}$ \cite{adesso20071}.

According to the ppt-criterion, a two-mode Gaussian state is separable if and only if the symplectic eigenvalues $\left\lbrace \tilde{\nu}_{1},\tilde{\nu}_{2}\right\rbrace $ of the partial transpose of the covariance matrix $\tilde{\vect V}_{j}$ with respect to the bipartition $j$ satisfy \cite{serafini20061,adesso20071}
\begin{equation}
\tilde{\nu}_{1},\tilde{\nu}_{2}\geq \frac{1}{2}.
\label{ppt}
\end{equation}
These are directly obtained from the roots $\left\lbrace \pm i\tilde{\nu}_{1},\pm i\tilde{\nu}_{2}\right\rbrace $ of the characteristic polynomial of the matrix $ \vect J^{T}_{2}\tilde{\vect V}_{j}$, which is given by
\begin{equation}
\lambda^{4}+\tilde{\Delta}^{2}_{1}\lambda^2+\tilde{\Delta}^{2}_{2}=0
\label{twomodeCP1}
\end{equation}
with $\tilde{\Delta}^{2}_{1}=\frac{1}{4}(a^ 2+b^2-2cd) $, $\tilde{\Delta}^{2}_{2}=\frac{1}{16}(ab-c^ 2)(ab-d^2)$, which are the symplectic invariants.

On other hand, inequality (\ref{pptR}) in the two-mode case may be translated into the eigenvalue problem of the product matrix \cite{horn1985},
\begin{equation}
\vect Z_{1}=4\vect P_{1}\left( \vect \Sigma +\vect V\right)\vect P_{1} \vect J_{n}^{T}\left( \vect \Sigma^{-1} +\vect V^{-1}\right)^{-1}  \vect J_{n}
\label{ZEing}
\end{equation} 
such that, inequality (\ref{pptR}) is not violated as long as all of the eigenvalues $\smash{\left\lbrace  \lambda_{z}^{(i)} ; i=1,2,3,4\right\rbrace }$ of $\vect Z_{1}$ are non-negative, {\it i.e.} $\lambda_{z}^{(i)}\geq 1 \quad \forall i $. 

Using the standard form (\ref{APPTSF1}) and substituting $\vect \Sigma_{m}$ by the covariance matrix of a pure squeezed state (see Eq. \ref{Sqezs})), $\vect Z_{1}$ results in the matrix $\vect Z_{1}(r)$ defined in Eq. (\ref{APPTZ}) whose entries are given in terms of rational functions in the squeezing parameter $r$, as discussed in more detail in appendix \ref{APPT}.

In the limit of infinite squeezing in momentum ($r \to 0$), we find that $\vect Z_{1}$ (see Eq.(\ref{twomodeLP})) has $\lambda_{z}^{(1)}=\lambda_{z}^{(2)}=1$ as doubly-degenerate eigenvalue, and the other two are given by the characteristic polynomial
\begin{equation}
\left( \frac{\lambda_z}{4}\right) ^{2}-\tilde{\Delta}^{2}_{1}\left( \frac{\lambda_z}{4}\right) +\tilde{\Delta}^{2}_{2} =0\ .
\label{twomodCP2}
\end{equation}
Since the roots of Eq.~\eqref{twomodCP2} are related with the roots of Eq.~\eqref{twomodeCP1} through the expression $\lambda=\pm i \sqrt{\lambda_z}/2$, the conditions $\lambda_z^{(3)}\ge 1$ and $\lambda_z^{(4)}\ge 1$ are indeed equivalent to Eq.~\eqref{ppt}. That is, given the optimal choice of probe states with $\ket{\varphi_{m}}=\ket{\varphi_{n+m}} $ ($m=1,...,n$) and infinitely-squeezed covariance matrix, we recover exactly the necessary and sufficient PPT criterion from the inequality (\ref{pptR}). It is straightforwardly to show that this assertion also holds if we consider infinite squeezing in position ($r \to \infty$) (see Eq.(\ref{twomodeLX})).

%%%%%%%%%%%%%%%%%%%%%%%%%%%%%%%%%%%%%%%%%%%%%%%%%%%%%%%%%%%%%%%%%%
\subsubsection{Three-mode Case}
The foregoing discussion sets the stage of the procedure that one has to follow in order to show the analogue result for pure three-mode Gaussian states. In this case, the comparison between the inequalities (\ref{pptR}) and (\ref{ppt}) has to be in terms of the three possible bipartitions of the system, such that the characteristic polynomial of the matrices $\vect Z_{j}$ ($j=1,2,3$) leads to the characteristic polynomial of the matrices $\vect J_{3}^{T} \tilde{\vect V}_{j}$. We defer the details of the proof to the appendix \ref{APPT}.

We may apply the same procedure to study the case of mixed tripartite-entangled states, but one finds that this assertion is not longer true. For mixed three-mode Gaussian states inequality (\ref{pptR}) can not be expected to reproduce the PPT criterion, since PPT basically discerns fully inseparability in the case of mixed states \cite{giedke20011,shalm20131}, whereas $\tau_{3,3}$ identifies genuine tri-partite entanglement. However, we found that $\tau_{2,3}$ still detects entanglement of the vast majority of three-mode bipartite entangled states.

%%%%%%%%%%%%%%%%%%%%%%%%%%%%%%%%%%%%%%%%%%%%%%%%%%%%%%%%%%%%%%%%%%%
\section{Examples}\label{SExa}

We now turn the attention to illustrate how expression (\ref{wholetauNG}) provides reliable estimates of $k$-partite entanglement in Gaussian and non-Gaussian states.

\begin{figure}[ht!]
\includegraphics[width=0.95\columnwidth]{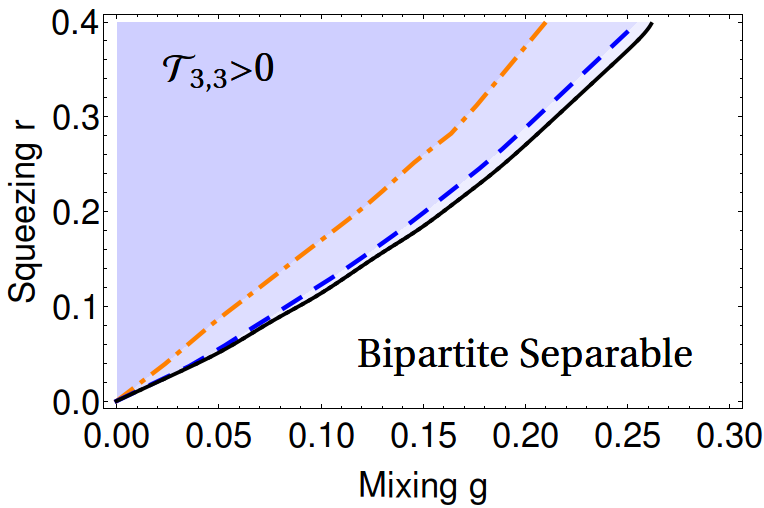} 
\caption{(color online). Density-map of the inseparability properties of the Werner-type GHZ state defined in Eq. (\ref{GHZm}) in terms of the mixing $g$ and squeezing parameter $r$. The black-solid line depicts the border between bipartite entangled (blue region) and separable states according to the PPT criterion. Within the former, the blue-dashed and orange-dot-dashed lines delimit the region of the states for which the hierarchies ${\cal T}_{2,3}$ and ${\cal T}_{3,3}$ return positive values, respectively.    \label{Fig1}}
\end{figure}
%%%%%%%%%%%%%%%%%%%%%%%%%%%%%%%%%%%%%%%%%%%%%%%%%%%%%%%%%%%%%%%%%%%
\subsection{Mixed genuine tripartite entangled states}

Let us start analyzing the inseparability properties of a mixed tripartite Gaussian entangled state, whose covariance matrix may be expressed as follows,
\begin{equation}
\vect V=\vect V_{GHZ}+g \vect I_{3}, \quad \text{with }  g\geq0,
\label{GHZm}
\end{equation}
where $\vect I_{n}=\bigoplus_{m=1}^{n}\vect I$, and
\begin{equation}
\vect V_{GHZ}=\frac{1}{2}\left( \begin{array}{cccccc}
    a      &     0     &   -c    &       0        &  -c  & 0        \\
    0          &   b   &   0             &   c   &       0      &  c  \\
    -c &   0       &     a       &   0            &     -c&   0    \\
    0          &  c&       0        &  b         &       0       &  c   \\
    -c &   0        &   -c   &   0           &       a    &  0\\
    0          &  c&       0       &  c    &       0       &  b \\
\end{array}\right),
\label{AEvolWCV} 
\end{equation}
with
\begin{eqnarray}
a&=& \frac{1}{2}\left(e^{2r} +\cosh(2r)\right) \nonumber \\
b&=& \frac{1}{2}\left(e^{-2r} +\cosh(2r)\right) \nonumber \\
c&=&\frac{1}{2}\sinh(2r) \nonumber ,
\end{eqnarray}
is the covariance matrix of the continuous-variable analogue of the GHZ states \cite{giedke20011}. Here, $g$ plays the role of a mixing parameter, while $r\geq 0$ is the squeezing parameter. We compare the hierarchies $\tau_{2,3}$ and $\tau_{3,3}$ with the PPT criterion applied to the bipartition $1|23$ \cite{giedke20011}.

As one can see in Fig. (\ref{Fig1}), $\tau_{3,3}$ detects that this state is genuinely tri-partite entangled in a substantial part in the parameter regime, and for sufficiently strong squeezing, even substantially mixed states are still genuinely tri-partite entangled. States that are too strongly mixed to be genuinely tri-partite entangled can still be identified to be bipartite entangled via $\tau_{2,3}$, which detects nearly as many states as the ppt criterion.

\begin{figure}[ht!]
\includegraphics[width=0.9\columnwidth]{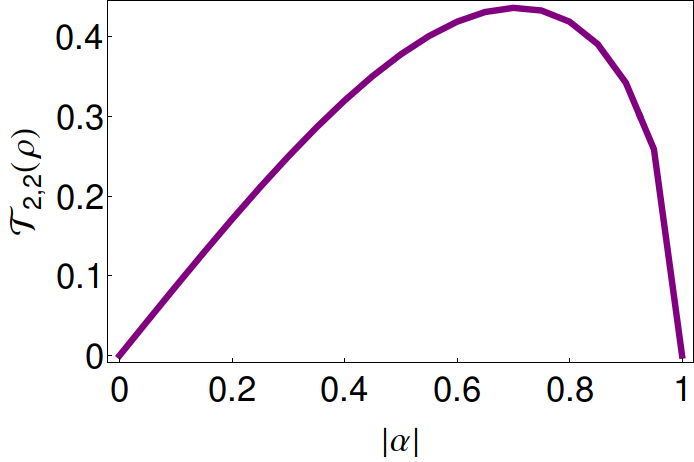} 
\caption{(color online). $ {\cal T}_{2,2}(\hat{\rho})$ as a function of the amplitude $|\alpha|$ for the CPS-TSVS defined in Eq.(\ref{CFx}) with $r=0$, $\alpha=|\alpha|e^{i\frac{\sqrt{2}}{2}}$, and $\beta=|\beta|e^{i\frac{\pi}{2}}$. \label{Fig2}}
\end{figure}
%%%%%%%%%%%%%%%%%%%%%%%%%%%%%%%%%%%%%%%%%%%%%%%%%%%%%%%%%%%%%%%%%%%%
\subsection{Coherent-Photon Added/Subtracted Two-mode States}

To demonstrate the performance on non-Gaussian states we investigate the inseparability properties of coherently photon-subtracted two mode squeezed vacuum states (CPS-TSVS). These states derive from the locally squeezed two-mode vacuum state by applying the operator $(\alpha \hat{a}_{1}+\beta\hat{a}_{2})^{u}$, where $\hat{a}_{l}$ ($l=1,2$) is the photon-annihilation operator of the $l$th mode and $|\alpha |^2+|\beta |^2=1 $ \cite{guo20131}. For simplicity, we shall consider the states obtained for $u=1$ and symmetrically squeezed in both modes. The covariance matrix $\vect V$ and the polynomial function $F$ that define the Wigner function via Eq. (\ref{WignerfG}) take the form, $\vect V=\frac{1}{2}diag(e^{-2r},e^{2r},e^{-2r},e^{2r})$, and
\begin{eqnarray}
F(\vect x)&=&2\cosh^2(r)\Big( (x_{1}^2+ p_{1}^2)|\alpha |^2+( x_{2}^2+ p_{2}^2)|\beta |^2 \nonumber \\
&+&2 \Re((x_{1}-i p_{1})(x_{2}+i p_{2})\alpha^{*}\beta)\Big)  \nonumber \\
&+&2\sinh^2(r)\Big( ( x_{1}^2+ p_{1}^2)|\alpha |^2+( x_{2}^2+ p_{2}^2)|\beta |^2 \nonumber 	\\
&+&2 \Re((x_{1}+i p_{1})(x_{2}-i p_{2})\alpha^{*}\beta)\Big) \nonumber \\
&-&4\cosh(r) \sinh(r)\Big( |\alpha p_{1}+\beta p_{2} |^2 \nonumber \\
&-&|\alpha x_{1}+\beta x_{2} |^2\Big) -1.
\label{CFx}
\end{eqnarray}
In \cite{agarwal20051} it is shown that the PPT criterion based on the second-order correlations fails to unveil the entanglement of this state for $r=0$,  what makes this state particularly interesting  to demonstrate the strength of the hierarchy. Remarkably enough, figure (\ref{Fig2}) shows that expression (\ref{wholetauNG}) is able to detect this purely non-Gaussian entanglement in agreement with \cite{agarwal20051}. Fig.(\ref{Fig2}) corresponds to a specific choice of the phases of the complex parameters $\alpha$ and $\beta$, but, we found $\tau_{2,2}$ to perform equally well for any other choice of phases.

\begin{figure}[ht!]
\includegraphics[width=0.9\columnwidth]{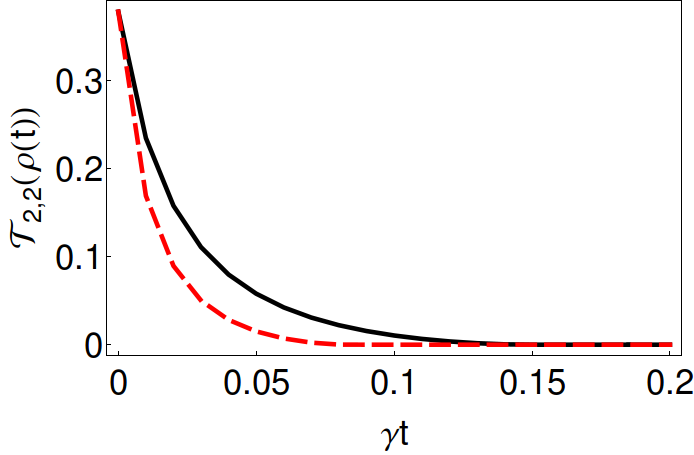} 
\caption{(color online). Time evolution of $ {\cal T}_{2,2}(\hat{\rho}(t))$ when the system is initially in the CPS-TSVS state plotted in Fig. (\ref{Fig2}) with $|\alpha|=0.5$, and it is in contact with independent thermal reservoirs with $N_{th}=2$ (black-solid line) and with $N_{th}=4$ (red-dashed line).}\label{Fig3}
\end{figure}

%%%%%%%%%%%%%%%%%%%%%%%%%%%%%%%%%%%%%%%%%%%%%%%%%%%%%%%%%%%%%%%%%%%
\subsection{Time evolution of an initially non-Gaussian entangled state}\label{SEAS}
Finally, the tractable form of the hierarchy (\ref{wholetauNG}) also permits to study the time evolution of the $k$-partite entanglement under the influence of environmental noise. Let us investigate how the two-mode non-Gaussian entanglement of the foregoing example is influenced when each mode is in contact with an independent heat bath. To be specific we assume the environmental coupling of both modes to be modelled with the same rate $\gamma$, and both baths to have the same temperature characterized by the mean photon number $N_{th}$. The open system dynamics is governed by a Fokker-Plank equation in the interaction picture (see Eq.\ref{Fokker-PlankE} in appendix \ref{AEVO}), which has been extensively employed to study the effects of losses and thermal hopping in CV systems \cite{serafini20041}.

The time-dependent Wigner function is obtained from the Green function of the Fokker-Plank equation (see appendix \ref{AEVO} for further details). In the interaction picture, one finds that the covariance matrix evolves according to
\begin{equation}
\vect V(t)=\vect {\varepsilon}(t)+\vect \sigma(t),
\label{FCS1}
\end{equation}
with
\begin{eqnarray}
\vect {\varepsilon}(t)&=& \frac{e^{-\gamma t}}{2}\vect V(0), \nonumber \\
\vect {\sigma}(t)&=&(1-e^{-\gamma t})\vect V_{(N_{th},0)}, \nonumber
\end{eqnarray}
where $\vect V(0)=\frac{1}{2}diag(e^{-2r},e^{2r},e^{-2r},e^{2r})$, $\vect V_{(N_{th},0)}=\frac{1+2N_{th}}{2}\left( \vect I\oplus \vect I\right) $, and the polynomial part $F(\vect x, t)$ is given by
\begin{eqnarray}
F(\vect x, t)&=&F\left(e^{\frac{\gamma}{2}t}(\vect \varepsilon^{-1}(t)\vect \sigma(t)+\vect I_{2})^{-1}\vect x \right) \nonumber \\
&+&\frac{1}{2}\sum_{l,m}\left( \vect \varepsilon^{-1}(t)+\vect \sigma^{-1}(t)\right)^{-1}_{lm}  \frac{\partial^{2}F(e^{\frac{\gamma}{2}t}\vect x)}{\partial\left[ \vect x\right] _{l}\partial\left[ \vect x\right] _{m}} \Bigg |_{\vect x=\vec{0}},
\label{FCS2}
\end{eqnarray}
For $t=0$, Eq.(\ref{FCS2}) returns the initial expression Eq.(\ref{CFx}) for the state ($F(\vect x,0)= F(\vect x)$), whereas in the long time ($F(\vect x,t\rightarrow \infty )\rightarrow 1$) the system evolves asymptotically into the symmetrical separable thermal (Gaussian) state.

One may appreciate from the figure \ref{Fig3} that the initial non-Gaussian entanglement is degraded \textit{asymptotically} in time: the hierarchies shows that the two-mode entanglement features an exponential decay. 

This example illustrates that Eq.(\ref{wholetauNG}) may provide an accurate description of multipartite CV entanglement in realistic dissipative scenarios. As the hierarchy deals with Gaussian and non-Gaussian states at the same footing, Eq.(\ref{wholetauNG}) is particularly of interested to study the time evolution of $k$-partite entanglement when the state evolves from Gaussian to non-Gaussian, or vice-versa.

%%%%%%%%%%%%%%%%%%%%%%%%%%%%%%%%%%%%%%%%%%%%%%%%%%%%%%%%%%%%%%%%%%%
\section{Experimental quantification}\label{SEAS}
Let us now briefly discuss how the hierarchies (\ref{wholetauNG}) and (\ref{Stau}) can be assessed with experimental data. The standard procedure would be based on the experimental reconstruction of the Wigner function in terms of quantum state tomography \cite{rehacek20091,lvovsky20091} or a measurement scheme specially designed for multicomponent CV systems \cite{tufarelli20121}, followed by the analytical evaluation of Eqs.(\ref{wholetauNG}) and (\ref{Stau}). However, the hierarchies for Gaussian states (\ref{Stau}) may be also directly accessed by performing Gaussian measurements, modelled in terms of a positive-valued operators with Gaussian Weyl symbol \cite{eisert20031,olivares20121}, which will be characterized by a covariance matrix $\vect \sigma_{M}$ and first-moment vector $\vect X_{M}$ that plays the role of the outcome of the measurement. If one performs such a measurement on the whole $n$-mode system, the probability of the outcome $\vect X_{M}$ is given by \cite{olivares20121}
\begin{equation}
p(\vect X_{M}; \vect \sigma_{M} )=\frac{e^{-\frac{1}{2}\vect X_{M}^{T}\frac{1}{\vect \sigma_{M} +\vect V}\vect X_{M} }}{(2\pi)^{n}\sqrt{\det\left(\vect \sigma_{M} +\vect V \right) }}.
\nonumber
\end{equation}
One may immediately identify the terms in the sum in Eq.(\ref{Stau}) as $(2\pi)^{n}p(\vect P_{j}\vect X; \vect \Sigma )$, since these terms are derived from diagonal matrix elements (see Eq.(\ref{ME2}), (\ref{ME3})). On the other hand, the first term in Eq.(\ref{Stau}), which results from off-diagonal matrix elements (see Eq.(\ref{ME1})), may be expressed in terms of the Fourier transform 
$\hat{p}(\vect \omega;\vect \Sigma)$ of the probability distribution $p(\vect X; \vect \Sigma)$,{ \it i.e.}
\begin{equation}
\hat{p}(\vect \omega; \vect \Sigma)=\frac{1}{(2\pi)^{n}} \int_{\mathbb{R}^{2n}}d^{2n}\vect X e^{-i\vect \omega^{T} \vect X }p(\vect X; \vect \Sigma),
\nonumber
\end{equation}
such that Eq.(\ref{Stau}) may be brought in the form,
\begin{eqnarray}
\tilde{\tau}_{k,n}(\hat{\rho})&=& e^{ -2\vect X^{T}\vect J_{n}^{T}\vect \Sigma \vect J_{n}\vect X } \int_{\mathbb{R}^{2n}}d^{2n}\vect \omega e^{-2\vect \omega^{T} \vect \Sigma \vect J_{n} \vect X } \hat{p}(\vect \omega; \vect \Sigma) \nonumber \\
&-&(2\pi)^{n}\sum_{j}a^{(k,n)}_{j}p(\vect P_{j}\vect X; \vect \Sigma ).
\label{StauE}
\end{eqnarray}
as we extensively show in appendix \ref{AEAS}. This expression relates $\tilde{\tau}_{k,n}$ directly to the measurement statistics of a Gaussian measurement with covariance matrix $ \vect \Sigma$.

Since the projection of $\hat{\rho}$ onto a one-mode pure infinitely-squeezed state (whose covariance matrix we illustrate in (\ref{Sqezs})) models an (ideal) homodyne measure in the $m$-th mode of the system \cite{eisert20021,giedke20021,lvovsky20091}, the results of Sec. \ref{SRPPT}   indicate that one may \textit{completely} certify the inseparability of arbitrary two-mode and pure three-mode Gaussian states by a collective of simultaneous (ideal) homodyne measures on each mode of the system.

%%%%%%%%%%%%%%%%%%%%%%%%%%%%%%%%%%%%%%%%%%%%%%%%%%%%%%%%%%%%%%%%%%%
\section{Concluding remarks and outlook}

The strength of the hierarchy as demonstrated by the explicit examples in Sec \ref{SExa} and the prospect to obtain a fine-grained characterization of multi-mode entanglement properties even for non-Gaussian states based only on Gaussian measurements underlines the practical value of the separability criteria presented here. In particular, the recent development of opto-mechanical experiments \cite{groblacher20091,brawley20141} that permit the realization of controlled interactions between massive degrees of freedom \cite{aspelmeyer20131} and light call for tools that permit to verify experimental achievements.
Whereas experiments on continuous variable entangled systems were in the realm of Gaussian states for a long time, this new generation of experiments permits to realize sizeable non-linear interactions which result in the generation of non-Gaussian entangled states.

This prospect to create and probe entangled states that were out of reach until recently, highlights the demand for theoretical tools for the analysis of entanglement properties beyond the Gaussian theory. In particular with the capacity to probe entanglement properties also in multi-mode systems, the present separability criteria promise to be a valuable theoretical support for a series of experiments to come.

\acknowledgments
The authors acknowledge useful discussions with D. Alonso, and \L . Rudnicki. A.A.V would like to thank D. Alonso and S. Kohler for their wise advices, and he is indebted to C. Dittrich, J. P\"{a}hle, and the group of "Coherent many-body quantum dynamics" at the Freiburg Institute for Advanced Studies their warm hospitality throughout his visit in Freiburg.  Financial support by the European Research Council under the project Odycquent is gratefully acknowledged. A.A.V. acknowledges financial support by the Government of the Canary Islands through an ACIISI fellowship (85\%co-financed by the European Social Fund), and by Vicerectorado de Investigaci\'{o}n de la Universidad de La Laguna.

\appendix

%%%%%%%%%%%%%%%%%%%%%%%%%%%%%%%%%%%%%%%%%%%%%%%%%%%%%%%%%%%%%%%%%%
\section{Derivation of Eq.(\ref{wholetauNG})}\label{ATool}
In this appendix we illustrate the derivation of expression (\ref{wholetauNG}) starting from the formulation Eq.(\ref{Tau}) of the hierarchy $\tau_{k,n}(\hat{\rho})$ in $\mathcal{H}_{n} $. The latter involves the following three matrix elements
\begin{eqnarray}
&\Braket{\Phi_{1}| \hat{\rho}|\Phi_{2}}& \label{ME1}, \\
&\Braket{\Phi_{1j}| \hat{\rho} |\Phi_{1j} } &\label{ME2},  \\
&\Braket{\Phi_{2j} | \hat{\rho} |\Phi_{2j} }& \label{ME3}. 
\end{eqnarray}
with $\ket{\Phi_{1}}$, $\ket{\Phi_{2}}$, $\ket{\Phi_{1j}}$, and $\ket{\Phi_{2j}}$ defined in Eqs. (\ref{dummy1}) and (\ref{dummy2}).
One may compute these matrix elements by using the trace product rule \cite{schleich2001},
\begin{eqnarray}
\Braket{ \vect \phi \left| \hat{\rho}  \right|  \vect  \psi } &=&\mbox{Tr} \left(   \hat{\rho} \left| \vect\psi \rangle\langle  \vect \phi\right|\right)   \nonumber \\
&=&(2\pi)^{n}\int d^{2n}\vect x W(\vect x)W_{\left| \vect\psi \rangle\langle  \vect \phi \right|} (\vect x).
\label{ATraceR}
\end{eqnarray}
Hence, we must first derive the Weyl symbol $W_{\ket{\Phi_2}\bra{\Phi_1}}$ corresponding to the $n$-fold tensor product operator $\left| \Phi_{2} \rangle\langle \Phi_{1} \right|=\bigotimes_{m=1}^{n}\ket{\varphi_{n+m}}\Bra{\varphi_{m}}$. According to the definition in Eq.(\ref{Wsymbol}), this may be expressed as
\begin{equation}
W_{\ket{\Phi_2}\bra{\Phi_1}}(\vect x)=\prod_{m=1}^{n}W_{\ket{\varphi_{n+m}}\bra{\varphi_{m}}}(q,p).
\label{AWsymbol1}
\end{equation}
Moreover, $ W_{\ket{\varphi_{n+m}}\bra{\varphi_{m}}}$ may be directly derived by using the classical formulation of the Wigner function \cite{schleich2001}, and the expression for the wave function of any single-mode pure Gaussian state, {\it i.e.}
\begin{equation}
\phi_{m}(q)=\sqrt{\frac{2 \sigma_{pp}^{(m)}}{\pi(1+4(\sigma_{xp}^{(m)})^2)}}e^{-\frac{\sigma_{pp}^{(m)}(q-\bar{q}_{m})^2}{1+2i\sigma_{xp}^{(m)}}+iq\bar{p}_{m}}.
\nonumber
\end{equation}
Doing so, one arrives at the Gaussian function
\begin{equation}
W_{\ket{\varphi_{l}}\bra{\varphi_{m}}}(q,p)=N_{m,l}e^{-\frac{1}{2}(((q,p)-\vect X_{m,l})^{T}\vect \Sigma_{m,l}^{-1}((q,p)-\vect X_{m,l})) } 
\label{AWsymbol}
\end{equation}
with first-moment $\vect X_{m,l}=1/2((\bar{q}_{m},\bar{p}_{m})+(\bar{q}_{l},\bar{p}_{l}))^{T}+i \vect \Sigma_{m,l}\vect J_{1}((\bar{q}_{m},\bar{p}_{m})-(\bar{q}_{l},\bar{p}_{l}))^{T} $ and covariance matrix as given in Eq.(\ref{CV3}), where the absolute value of the normalizing factor is given by
\begin{equation}
|N_{m,l}|=\dfrac{e^{\frac{-1}{4\det(\vect \Sigma_{m}+\vect \Sigma_{l})}\left((\vect X^{-})^{T}\vect J_{1}^{T}(\vect \Sigma_{m}+\vect \Sigma_{l} ) \vect J_{1}\vect X^{-}  \right)} }{\pi \sqrt[4]{\det(\vect \Sigma_{m}+\vect \Sigma_{l})}},
\nonumber
\end{equation}
with $\vect X^{-}=(\bar{q}_{m},\bar{p}_{m})-(\bar{q}_{l},\bar{p}_{l})$. Notice that, from Eq.(\ref{CV3}) it is deduced that $\vect \Sigma_{\Phi_{21}}$ is a complex symmetric matrix which in general is not Hermitian. One may follow the same recipe to obtain the other Weyl symbols corresponding to the operators $ \left| \Phi_{1j} \rangle\langle \Phi_{1j} \right|$, and $ \left| \Phi_{2j} \rangle\langle \Phi_{2j} \right|$.

By virtue of the trace product rule (\ref{ATraceR}), the matrix element (\ref{ME1}) takes the form,
\begin{widetext}
\begin{eqnarray}
&&\Braket{ \Phi_{1} | \hat{\rho} |   \Phi_{2}} = \frac{N_{\Phi_{21}}}{\sqrt{\det(\vect V)}}\int_{\mathbb{R}^{2n}}d^{2n}\vect x \ F(\vect x)e^{-\frac{1}{2}\vect x^{T} \vect V^{-1} \vect x}e^{-\frac{1}{2}(\vect x-\vect X_{\Phi_{21}})^{T}\vect \Sigma_{\Phi_{21}}^{-1}(\vect x-\vect X_{\Phi_{21}})} \nonumber \\
&=&\frac{N_{\Phi_{21}}}{\sqrt{\det(\vect V)}}\int_{\mathbb{R}^{2n}}d^{2n}\vect x \left( F(\vect x) e^{\frac{1}{2}\left( \vect x^{T}\vect \Sigma_{\Phi_{21}}^{-1} \vect X_{\Phi_{21}} +\vect X_{\Phi_{21}}^{T}\vect \Sigma_{\Phi_{21}}^{-1}\vect x-\vect X_{\Phi_{21}}^{T}\vect \Sigma_{\Phi_{21}}^{-1}\vect X_{\Phi_{21}}\right) }\right) e^{-\frac{1}{2}\vect x^{T}\left(  \vect V^{-1}+\vect \Sigma_{\Phi_{21}}^{-1} \right) \vect x}\nonumber \\
&=&\frac{(2\pi)^{n} N_{\Phi_{21}} e^{-\frac{1}{2}\vect X_{\Phi_{21}}^{T}\vect \Sigma_{\Phi_{21}}^{-1}\vect X_{\Phi_{21}}}}{\sqrt{\det(\vect V^{-1}+\vect \Sigma_{\Phi_{21}}^{-1})\det(\vect V)}} \left[ e^{\frac{1}{2}\left( \frac{\partial}{\partial \vect x}\right)^{T} \left( \vect V^{-1}+\vect \Sigma_{\Phi_{21}}^{-1}\right)^{-1} \left( \frac{\partial}{\partial \vect x}\right) }\left( F(\vect x) e^{\vect X_{\Phi_{21}}^{T}\vect \Sigma_{\Phi_{21}}^{-1}\vect x }\right) \right]_{\vect x=\vec{0}} ,
 \nonumber
\end{eqnarray}
\end{widetext}
where we made use of the symmetry property of the pseudo-covariance matrix $\vect \Sigma_{\Phi_{21}}=\vect \Sigma_{\Phi_{21}}^T$. In this expression, $\vect x$ is $2n$-dimensional real vector. Since the exponential of the differential operator describes a shift in phase space (see appendix \ref{AEAS}), we can conveniently manipulate this expression to obtain,
\begin{widetext}
\begin{eqnarray}
\Braket{  \Phi_{2} | \hat{\rho} |   \Phi_{1}} =\frac{\pi^{n} N_{\Phi_{21}}e^{-\frac{1}{2}\vect X_{\Phi_{21}}^{T}\vect \Sigma_{\Phi_{21}}^{-1}\vect X_{\Phi_{21}}}}{\sqrt{\det(\vect V+\vect \Sigma_{\Phi_{21}})}}  \left[ e^{\frac{1}{2}\left( \frac{\partial}{\partial \vect x}+\vect \Sigma_{\Phi_{21}}^{-1} \vect X_{\Phi_{21}}\right)^{T} \left( \vect V^{-1}+\vect \Sigma_{\Phi_{21}}^{-1}\right)^{-1} \left( \frac{\partial}{\partial \vect x}+\vect \Sigma_{\Phi_{21}}^{-1} \vect X_{\Phi_{21}}\right) } F(\vect x) \right]_{\vect x=\vec{0}}.
\label{IntNG}
\end{eqnarray}
\end{widetext}
Similarly, one may derive the analogue expression for the matrix elements given in Eqs. (\ref{ME2}) (\ref{ME3}) by substituting the pair $\vect X_{\Phi_{21}}$, $\vect \Sigma_{\Phi_{21}}$ for the corresponding pair $\vect X_{\Phi_{1j} }$, $\vect \Sigma_{\Phi_{1j} }$, and  $\vect X_{\Phi_{2j} }$, $\vect \Sigma_{\Phi_{2j} }$ in Eq. (\ref{IntNG}) (and by taking $N_{\Phi_{21}}$ equal to $\pi^{-n}$). After replacing the result for each matrix element in Eq.(\ref{Tau}) and some straightforward algebra, one arrives at expression Eq.(\ref{wholetauNG}) for the hierarchy that is valid as long as the Wigner function of the system can be expressed as in Eq.(\ref{WignerfG}).

%%%%%%%%%%%%%%%%%%%%%%%%%%%%%%%%%%%%%%%%%%%%%%%%%%%%%%%%%%%%%%%%%%
\section{First-moment vectors and covariance matrices associated to the bipartition $j$}\label{APerm}
In this appendix we describe in more detail how to obtain the vectors $\vect X_{\Phi_{1j}}$ and $\vect X_{\Phi_{2j}}$, and the matrices $\vect \Sigma_{\Phi_{1j}}$, $\vect\Sigma_{\Phi_{2j}} $ and $\vect P_{j}$ associated with the bipartition labelled by $j$. In Sec. \ref{SHIC}, we stated that $\ket{\Phi_{1j}}$ and $\ket{\Phi_{2j}}$ are obtained from $\ket{\Phi_{1}}$ and $\ket{\Phi_{2}}$ by interchanging the one-mode states $\ket{\varphi_{m}}$ with $\ket{\varphi_{m+n}}$ corresponding to those subsystems that are grouped together in the bipartition $j$ (see Eq.(\ref{dummy2})). On the other hand, from Eqs. (\ref{X1}) and (\ref{CV3}) one obtains that the first-moment vectors of $\ket{\Phi_{1}}$ and $\ket{\Phi_{2}}$ are given by,
\begin{eqnarray}
\vect X_{\Phi_{1}}&=&\bigoplus^{n}_{m=1}\bar{\vect x}_{m},   \label{Xm1} \\
\vect X_{\Phi_{2}}&=&\bigoplus^{n}_{m=1}\bar{\vect x}_{n+m}.  \label{Xm2} 
\end{eqnarray}
and the covariance matrices are given by
\begin{eqnarray}
\vect \Sigma_{\Phi_{1}}&=&\bigoplus^{n}_{m=1}\vect \Sigma_{m},   \label{CV1} \\
\vect \Sigma_{\Phi_{2}}&=&\bigoplus^{n}_{m=1}\vect \Sigma_{n+m}.  \label{CV2} 
\end{eqnarray}
Analogously, one may deduce the covariance matrices $\vect \Sigma_{\Phi_{1j}}$ and $\vect\Sigma_{\Phi_{2j}} $ by permuting the corresponding matrices $\vect \Sigma_{m}$ and $\vect\Sigma_{n+m} $ in the expressions (\ref{CV1}) and (\ref{CV2}), respectively. Doing so, one obtains that,
\begin{eqnarray}
\vect \Sigma_{\Phi_{1j}}&=&\bigoplus_{m=1}^{n}\vect \Sigma_{m+n \left[ \vect v_{j}\right] _{m}} \ ,  \label{CV4} \\  
\vect \Sigma_{\Phi_{2j}}&=&\bigoplus_{m=1}^{n}\vect \Sigma_{m+n -n\left[ \vect v_{j}\right] _{m}} \ .\label{CV5}
\end{eqnarray}

The same reasoning may be applied to derive the first-moment vectors, where one interchanges the corresponding vectors $ \bar{\vect x}_{m}$ and $\bar{ \vect x}_{n+m} $ in Eqs.(\ref{Xm1}) and (\ref{Xm2}). These permutations may be expressed in a compact way with the matrix $P$ defined in Eq.(\ref{Pmatrix}), such that $\vect X_{\Phi_{1j}}$ and $\vect X_{\Phi_{2j}}$ may be written as \cite{mintert20091} 
\begin{eqnarray}
\vect X_{\Phi_{1j}}&=& \frac{\vect X_{\Phi_{1}}+\vect X_{\Phi_{2}}}{2} + \frac{1}{2}\vect P_{j}(\vect X_{\Phi_{1}}-\vect X_{\Phi_{2}}),  \label{X2}   \\
\vect X_{\Phi_{2j}}&=&\frac{\vect X_{\Phi_{1}}+\vect X_{\Phi_{2}}}{2} -\frac{1}{2} \vect  P_{j}(\vect X_{\Phi_{1}}-\vect X_{\Phi_{2}}). \label{X3}   
\end{eqnarray}

%%%%%%%%%%%%%%%%%%%%%%%%%%%%%%%%%%%%%%%%%%%%%%%%%%%%%%%%%%%%%%%%%%%%%
\section{Resemblance to the PPT Criterion}\label{APPT}

%%%%%%%%%%%%%%%%%%%%%%%%%%%%%%%%%%%%%%%%%%%%%%%%%%%%%%%%%%%%%%%%%%%%%%%
\subsection{Two-mode Gaussian case}
The standard form of the covariance matrix of any two-mode Gaussian state reads \cite{adesso20071}
\begin{equation}
\vect V=\frac{1}{2}\left( \begin{array}{cccc}
    a &  0   &   c   &   0\\
    0&   a   &   0   &   d\\
    c&   0   &   b   &   0\\
    0&   d   &   0   &   b\\
\end{array}\right) ,
\quad \left\lbrace a,b,c,d \right\rbrace \in \mathbb{R}^4, 
\label{APPTSF1} 
\end{equation} 
whereas the covariance matrix of a one-mode pure squeezed state may be expressed as follows
\begin{eqnarray}
\vect \Sigma(r)&=&diag\left( \frac{1}{4r},r\right),
\label{Sqezs}
\end{eqnarray}
where $r$ is the squeezing parameter.

After substituting Eqs. (\ref{APPTSF1}) and (\ref{Sqezs}) in the expression for the matrix (\ref{ZEing}), one obtains that the latter takes the following form
\begin{widetext}
\begin{equation}
\vect Z_{1}(r)=\left( \begin{array}{cccc}
    \frac{(1+2ar)(a(b+2r)-d^2)-4cdr^2 }{(a+2r)(b+2r)-d^2} &  0 & \frac{2r(2r(ad-bc)+d(1+cd)-abc)}{(a+2r)(b+2r)-d^2}    & 0 \\
   0&    \frac{(a+2r)(a(1+2rb)-2rc^2)-cd}{(1+2ar)(1+2br)-4c^2r^2}&  0     & \frac{2r(c(1+cd)-abd)+ac-bd}{(1+2ar)(1+2br)-4c^2r^2}   \\  
  \frac{2r(2r(bd-ac)+d(1+cd)-abc) }{(a+2r)(b+ar)-d^2}   &      0        &    \frac{(1+2br)(b(a+2r)-d^2)-4cdr^2}{(a+2r)(b+2r)-d^2}          & 0   \\
   0 &      \frac{2r(c(1+cd)-abd)+bc-ad}{(1+2ar)(1+2br)-4c^2r^2}        &    0     &       \frac{(b+2r)(b(1+2ra)-2rc^2)-cd}{(1+2ar)(1+2br)-4c^2r^2}        \\    
\end{array}\right) .
\label{APPTZ}
\end{equation}
\end{widetext}
As one may see, the entries of the matrix $\vect Z_{1}(r)$ are rational functions in terms of the squeezing parameter $r$, and the limit $r\rightarrow 0$ reads
\begin{equation}
\lim_{r\rightarrow 0}\vect Z_{1}(r)=\left( \begin{array}{cccc}
    1 &  0   &   0   &   0\\
    0&   a^2-cd   &   0   &   ac-bd\\
    0&   0   &  1   &   0\\
    0&   bc-ad   &   0   &   b^2-cd\\
\end{array}\right) .
\label{twomodeLP}
\end{equation}
Similarly, one may derive the expression for $\vect Z_{1}(r)$  in the limit $r\rightarrow \infty$, which corresponds to an infinite squeezing in position. Doing so, one may replace $r$ in (\ref{APPTZ}) by $1/r$, and then take the limit $r\rightarrow 0$, {\it i.e.}
\begin{equation}
\lim_{r\rightarrow 0}\vect Z_{1}\left( \frac{1}{r}\right) =\left( \begin{array}{cccc}
     a^2-cd &  0   &   -bc+ad   &   0\\
       0    &  1   &   0        &  0  \\
    -ac+bd  &  0   &    b^2-cd  &   0  \\
       0    &   0   &   0        &  1\\
\end{array}\right) .
\label{twomodeLX}
\end{equation}
Both (\ref{twomodeLP}) and (\ref{twomodeLX}) have $\lambda^{(1)}_{z}=\lambda^{(1)}_{z}=1$ as a doubly degenerate eigenvalue. The other two eigenvalues are given by $\lambda^{(3)}_{z}=4 \tilde{\nu}_{1}^2$ and $\lambda^{(3)}_{z}=4\tilde{\nu}_{1}^2$, as we point out in Sec. \ref{SRPPT}. This illustrates that the hierarchy expressed in terms of the inequality (\ref{pptR}) reproduces the results of the PPT criterion when we choose infinitely-squeezed probe states either in momentum or position.

%%%%%%%%%%%%%%%%%%%%%%%%%%%%%%%%%%%%%%%%%%%%%%%%%%%%%%%%%%%%%%%%%%%%%%%
\subsection{Three-mode Gaussian case}

The standard form of a pure three-mode Gaussian state reads \cite{adesso20071}
\begin{equation}
\vect V=\frac{1}{2}\left( \begin{array}{cccccc}
    a_{1}      &     0     &   e^{+}_{12}    &       0        &  e^{+}_{13}  & 0 \\
    0          &   a_{1}   &   0             &   e^{-}_{12}   &       0      &  e^{-}_{13}\\
    e^{+}_{12} &   0       &     a_{2}       &   0            &     e^{+}_{23}&   0\\
    0          &  e^{-}_{12}&       0        &  a_{2}         &       0       &   e^{-}_{23}\\
    e^{+}_{13} &   0        &   e^{+}_{23}   &   0            &       a_{3}    &  0\\
    0          &  e^{-}_{13}&       0        &  e^{-}_{23}    &       0       &  a_{3}\\
\end{array}\right),
\label{APPTSF2} 
\end{equation}
where $a_{1},a_{2},a_{3}\in \mathbb{R}$, and $ e^{\pm}_{12}$, $e^{\pm}_{13}$, $e^{\pm}_{23}$ are simple functions of $a_{1}$,$a_{2}$, and $a_{3}$.

The characteristic polynomial reads $\lambda^{6}+\tilde{\Delta}^{3}_{1}\lambda^{4}+ \tilde{\Delta}^{3}_{2} \lambda^{2}+\tilde{\Delta}^{3}_{3} =0 $, and the symplectic invariants $\left\lbrace \tilde{\Delta}_{l}^{3}\right\rbrace$ ($l=1,2,3$) are obtained from \cite{serafini20061} 
\begin{equation}
\tilde{\Delta}_{l}^{3}=M_{2l}(\vect J_{3}^{T}\tilde{\vect V}_{1}^{2}),
\nonumber
\end{equation}
where $M_{2l}(\vect J_{3}^{T}\tilde{\vect V}_{1}^{2})$ is the principal minor of order $2l$ of the matrix $\vect J_{3}^{T}\tilde{\vect V}_{1}^{2}$, {\it i.e.} it is the sum of all the determinants of all the $2l\times 2l$ submatrices obtained by deleting $6-2l$ rows and the corresponding $6-2l$ columns \cite{serafini20061}. Since one has to follow the same procedure for each bipartition, we illustrate here only the case for $S_{1}|S_{2}S_{3}$, where $S_{m}$ symbolizes the $m$-th mode ($m=1,2,3$). Although the whole expression of $\vect Z_{S_{1}|S_{2}S_{3}}(r)$ is straightforwardly derived from (\ref{ZEing}) by replacing $\vect \Sigma_{m}=\vect \Sigma(r)$ for $m=1,2,3$ (its entries are again rational functions in terms of the squeezing parameter $r$), it is rather lengthy so that we only provide the final expression after taking the limit $r\rightarrow 0$,
\begin{widetext}
\begin{equation}
\lim_{r\rightarrow 0} \vect Z_{S_{1}|S_{2}S_{3}}(r)=\left( \begin{array}{cccccc}
    1      &     0     &   0         &       0        &     0             & 0 \\
    0  &   a_{1}^2-e^{+}_{13} e^{-}_{13}-e^{+}_{12}e^{-}_{12}   & 0 &   a_{1}e^{+}_{12}-a_{2}e^{-}_{12}-e^{-}_{13}e^{+}_{23} & 0 &   a_{1}e^{+}_{13}-a_{3}e^{-}_{13}-e^{-}_{12}e^{+}_{23}\\
    0 &   0       &    1       &   0            &     0&   0\\
    0  &   a_{2}e^{+}_{12}-a_{1}e^{-}_{12}+e^{+}_{13}e^{-}_{23}   & 0 &   a_{2}^2-e^{+}_{12}e^{-}_{12}+e^{+}_{23}e^{-}_{23} & 0 &   a_{2}e^{+}_{23}+a_{3}e^{-}_{23}-e^{-}_{12}e^{+}_{13}\\
     0 &   0       &    0     &   0            &     1&   0\\
    0  &   a_{3}e^{+}_{13}-a_{1}e^{-}_{13}+e^{+}_{12}e^{-}_{23}   & 0 &   a_{3}e^{+}_{23}
    +a_{2}e^{-}_{23}-e^{-}_{13}e^{+}_{12} & 0 &   a_{3}^2-e^{+}_{13}e^{-}_{13}+e^{-}_{23}e^{+}_{23}\\
\end{array}\right) .
\label{threemodeLP} 
\end{equation}
\end{widetext}
This matrix has $\lambda_{z}^{(1)}=\lambda_{z}^{(2)}=\lambda_{z}^{(3)}=1$ as a three-times degenerate eigenvalue, and the other eigenvalues are the roots of the polynomial
\begin{equation}
-\left( \frac{\lambda_{z}}{4}\right)^{3}+\tilde{\Delta}^{3}_{1}\left( \frac{\lambda_{z}}{4}\right)^{2}- \tilde{\Delta}^{3}_{2}\left( \frac{\lambda_{z}}{4}\right)+\tilde{\Delta}^{3}_{3} =0.
\label{cpthreemode}
\end{equation}
As we have already seen for the two-mode case, the roots of the characteristic polynomial of $\vect J_{3}^{T}\tilde{\vect V}^{2}_{1}$ are related to those of (\ref{cpthreemode}) through the expression $\lambda= \pm i\sqrt{\lambda_{z}}/2$. Hence, the inequality (\ref{pptR}) applied in the bipartition $ S_{1}|S_{2}S_{3}$ reproduces the PPT criterion for pure three-mode Gaussian states.

Analogously, one may show that this assertion holds for the other bipartitions $S_{2}|S_{1}S_{3}$ and $S_{3}|S_{1}S_{2}$. Now the roots of the corresponding characteristic polynomial are $\left\lbrace 1,1,1,4 \tilde{\nu}_{S_{2}|S_{1}S_{3},1}^{2},4 \tilde{\nu}_{S_{2}|S_{1}S_{3},2}^{2},4 \tilde{\nu}_{S_{2}|S_{1}S_{3},3}^{2}\right\rbrace $ and $\left\lbrace 1,1,1,4 \tilde{\nu}_{S_{3}|S_{1}S_{2},1}^{2},4 \tilde{\nu}_{S_{3}|S_{1}S_{2},2}^{2},4 \tilde{\nu}_{S_{3}|S_{1}S_{2},3}^{2}\right\rbrace $, in terms of the symplectic eigenvalues of the partially transpose covariance matrix corresponding to the bipartitions $S_{2}|S_{1}S_{3}$ and $S_{3}|S_{1}S_{2}$, respectively.

Once again, it is important note that the assertion also holds for infinite squeezing in position ($r\rightarrow \infty$). One gets at the following matrix for $\vect Z_{1}(r)$, which is analogue to (\ref{threemodeLP}),
\begin{widetext}
\begin{equation}
\lim_{r\rightarrow 0}\vect Z_{S_{1}|S_{2}S_{3}}\left( \frac{1}{r}\right) =\left( \begin{array}{cccccc}
    a_{1}^2-e^{+}_{13} e^{-}_{13}-e^{+}_{12}e^{-}_{12}       &     0     &   a_{1}e^{-}_{12}-a_{2}e^{+}_{12}-e^{+}_{13}e^{-}_{23}     &       0        &    a_{1}e^{-}_{13}-a_{3}e^{+}_{13}-e^{+}_{12}e^{-}_{23}     & 0 \\
    0  &  1   & 0    & 0   & 0 &  0  \\
     a_{2}e^{-}_{12}-a_{1}e^{+}_{12}+e^{-}_{13}e^{+}_{23}       &    0     &   a_{2}^2-e^{+}_{12}e^{-}_{12}+e^{+}_{23}e^{-}_{23}  &      0           &     a_{2}e^{-}_{23}+a_{3}e^{+}_{23}-e^{+}_{12}e^{-}_{13}&   0\\
    0  &   0   & 0 &  1 & 0 &   0\\
   a_{3}e^{-}_{13}-a_{1}e^{+}_{13}+e^{-}_{12}e^{+}_{23}       &    0     &   a_{3}e^{-}_{23}  +a_{2}e^{+}_{23}-e^{+}_{13}e^{-}_{12}   & 0     &     a_{3}^2-e^{+}_{13}e^{-}_{13}+e^{-}_{23}e^{+}_{23}&   0\\
    0  &   0   & 0 &  0  & 0 &    1  \\
\end{array}\right) .
\label{threemodeLX}
\end{equation}
\end{widetext}
from which one obtains the same characteristic polynomial as given in (\ref{cpthreemode}).

%%%%%%%%%%%%%%%%%%%%%%%%%%%%%%%%%%%%%%%%%%%%%%%%%%%%%%%%%%%%%%%%%%%%%%%%%%%%
\section{Time evolution of the Wigner function}\label{AEVO}
We consider the time evolution of an $n$-mode system governed by the Fokker-Plank equation in the interaction picture \cite{risken19961,serafini20041}
\begin{eqnarray}
\frac{\partial W(\vect x,t)}{\partial t}=\left(  \left( \frac{\partial}{\partial\vect x} \right)^{T}\vect  \Gamma  \vect x+\left( \frac{\partial}{\partial\vect x} \right)^{T}\vect D\frac{\partial}{\partial\vect x}\right) W(\vect x,t)
\label{Fokker-PlankE}
\end{eqnarray}
with $\left(  \frac{\partial}{\partial\vect x} \right) ^{T}=\bigoplus_{l=1}^{n}\left(\frac{\partial}{\partial q_{l}},\frac{\partial}{\partial p_{l}} \right)  $; $\vect \Gamma $ and $\vect D$ are $2n\times 2n$ real symmetric matrices that encode the interaction with the environment. In the case of interest here, these take the form $\vect \Gamma=\gamma/2\left( \vect I\oplus \vect I\right) $ and $\vect D=\gamma(1+2N_{th})/4\left( \vect I\oplus \vect I\right)$, where $N_{th}$ is the mean photon number of the baths.

Eq.(\ref{Fokker-PlankE}) is a linear Fokker-Plank equation with time-independent coefficients that can be straightforwardly solved by using the Green function method \cite{carmichael20021}, that permit to relate $W(\vect x,t)$ and $W(\vect x,0)$ via
\begin{eqnarray}
W(\vect x,t)=\int_{\mathbb{R}^{2n}} d^{2n}\vect x' \ W(\vect x',0)G(\vect x, \vect x', t).
\label{IntGreenF}
\end{eqnarray}
in terms of the Green function $G(\vect x, \vect x', t)$ which takes the form (see \cite{agarwal19711,carmichael20021})
\begin{eqnarray}
G(\vect x, \vect x', t)=\frac{1}{(2\pi)^n \sqrt{\det(\vect \sigma(t))}}e^{-\frac{1}{2}(\vect x-\vect b(t)\vect x')^{T}\vect \sigma(t)^{-1}(\vect x-\vect b(t)\vect x')}
\label{GreenF}
\end{eqnarray}
where
\begin{eqnarray}
\vect b(t)&=& e^{ -\vect \Gamma t},  \nonumber \\
\vect \sigma(t) &=&\vect \sigma(\infty)- e^{ -\vect \Gamma t}\vect \sigma(\infty)e^{ -\vect \Gamma t},  \nonumber
\end{eqnarray}
and $\vect \sigma(\infty) $ is the stationary solution of Eq(\ref{Fokker-PlankE}), which is obtained from solving
\begin{equation}
\vect \Gamma\vect \sigma(\infty)+\vect \sigma(\infty)\vect \Gamma=2\vect D.
\nonumber
\end{equation}
The integration of expression (\ref{IntGreenF}) with the Wigner function $ W(\vect x,0)$ of CPS-TSVS state results in the solutions depicted in Eqs. (\ref{FCS1}) and (\ref{FCS2}).

%%%%%%%%%%%%%%%%%%%%%%%%%%%%%%%%%%%%%%%%%%%%%%%%%%%%%%%%%%%%%%%%%%%%%%%%%%%%%%%%%%
\section{Experimental quantification}\label{AEAS}
In this section we will show the derivation of the following identity
\begin{eqnarray}
&&\Braket{\Phi_{1}| \hat{\rho}|\Phi_{2}  }=(2\pi)^{n} \int_{\mathbb{R}^{2n}}d^{2n}\vect x  \ W(\vect x)W_{\ket{\Phi_2}\bra{\Phi_1}}(\vect x)  \nonumber \\
&=&e^{ -2\vect X^{T}\vect J_{n}^{T}\vect \Sigma \vect J_{n}\vect X } \int_{\mathbb{R}^{2n}}d^{2n}\vect \omega \ e^{-2\vect \omega^{T} \vect \Sigma \vect J_{n} \vect X } \hat{p}(\vect \omega; \vect \Sigma),
\label{AEA1}
\end{eqnarray}
which has been used to obtain Eq.(\ref{StauE}) of Sec. \ref{SEAS}. To start with, the probability distribution $p(\vect X;\vect \Sigma) $, corresponding a Gaussian measurement with covariance matrix $\vect \Sigma$ and first-moment vector $\vect X$ on an $n$-mode system with Wigner function $W(\vect x)$, is given by 
\begin{equation}
p(\vect X;\vect \Sigma)=\int_{\mathbb{R}^{2n}}d^{2n}\vect x \ W(\vect x)\frac{e^{-\frac{1}{2}(\vect x-\vect X)^{T}\vect \Sigma^{-1}(\vect x-\vect X)}}{(2\pi)^{n}\sqrt{\det(\vect \Sigma)}}.
\label{AEA2}
\end{equation}
Introducing an unitary transformation $\vect U$, such that $\vect D=\vect U^{T} \vect \Sigma \vect U $ (or $\vect D^{-1}=\vect U^{T} \vect \Sigma^{-1} \vect U $) is a diagonal matrix, permits to rephrase this as
\begin{equation}
p(\vect U \tilde{\vect X}; \vect D)=\int_{\mathbb{R}^{2n}}d^{2n}\tilde{\vect x} \ W(\vect U\tilde{\vect x})\frac{e^{-\frac{1}{2}(\tilde{\vect x}-\tilde{\vect X})^{T}\vect D^{-1}(\tilde{\vect x}-\tilde{\vect X})}}{(2\pi)^{n}\sqrt{\det(\vect D)}}
\nonumber
\end{equation}
with $\vect x=\vect U \tilde{\vect x} $ and  $\vect X=\vect U \tilde{\vect X} $, where we have used  $d^{2N}\tilde{\vect x} =d^{2N}\vect x$  since the Jacobian determinant $|\det(\vect U)|=1$. From here it becomes clear that $p(\vect X; \vect \Sigma)$ can be considered a multidimensional convolution transform with a Gaussian kernel, and its inverse formula is well established \cite{polyanin20071}. Using the latter for Eq.(\ref{AEA2}), one obtains
\begin{equation}
W(\vect X)= exp\left( -\frac{1}{2}\left( \frac{\partial}{\partial \vect X}\right)^{T} \vect \Sigma\left( \frac{\partial}{\partial \vect X}\right) \right)p(\vect X; \vect \Sigma ).
\nonumber
\end{equation}
We may derive a more suitable form for this expression by using the Fourier transform of the Dirac delta function, that is
\begin{widetext}
\begin{eqnarray}
W(\vect x)&=& exp\left( -\frac{1}{2}\left( \frac{\partial}{\partial \vect x}\right)^{T} \vect \Sigma\left( \frac{\partial}{\partial \vect x}\right) \right)\int_{\mathbb{R}^{2n}}d^{2n}\vect u \ p(\vect u;\vect \Sigma) \delta(\vect x-\vect u) \nonumber \\
&=&\frac{1}{(2\pi)^{2n}} \int_{\mathbb{R}^{2n}}\int_{\mathbb{R}^{2n}}d^{2n}\vect u d^{2n}\vect \omega \ p(\vect u;\vect \Sigma) exp\left( -\frac{1}{2}\left( \frac{\partial}{\partial \vect x}\right)^{T} \vect \Sigma\left( \frac{\partial}{\partial \vect x}\right) \right)exp\left(i\vect \omega^{T} (\vect x-\vect u) \right)  \nonumber \\
&=&\frac{1}{(2\pi)^{2n}} \int_{\mathbb{R}^{2n}}\int_{\mathbb{R}^{2n}}d^{2n}\vect u d^{2n}\vect \omega  \ p(\vect u;\vect \Sigma)exp\left( \frac{1}{2} \vect \omega^{T} \vect \Sigma \vect \omega  \right)exp\left(i\vect \omega^{T} (\vect x-\vect u) \right).
\label{AEA3}
\end{eqnarray}

On the other hand, the phase-space counterpart of $\left| \Phi_{2}  \rangle\langle \Phi_{1} \right|  $ is given by, 

\begin{equation}
W_{\left| \Phi_{2}  \rangle\langle \Phi_{1} \right| }(\vect x)=\frac{1}{(2\pi)^{n}\sqrt{\det(\vect \Sigma)}} e^{ -2\vect X^{T}\vect J_{n}^{T}\vect \Sigma \vect J_{n}\vect X }e^{ -\frac{1}{2}(\vect x-2i\vect \Sigma\vect J_{n} \vect X)^{T}\vect \Sigma^{-1} (\vect x-2i\vect \Sigma\vect J_{n}\vect X )},
\label{AEA4}
\end{equation}
according to Eqs. (\ref{AWsymbol1}) and (\ref{AWsymbol}) in appendix \ref{ATool}. By replacing Eq.(\ref{AEA3}) and (\ref{AEA4}) in the expression for the matrix element (\ref{AEA1}), one obtains
\begin{eqnarray}
\Braket{\Phi_{1}| \hat{\rho}|\Phi_{2}  } & =&
(2\pi)^{n} \int_{\mathbb{R}^{2n}}d^{2n}\vect x \ W(\vect x)W_{\ket{\Phi_2}\bra{\Phi_1}}(\vect x) \nonumber \\
&=&\frac{1}{(2\pi)^{n}}\int_{\mathbb{R}^{2n}}\int_{\mathbb{R}^{2n}}d^{2n}\vect \omega d^{2n}\vect u   \ p(\vect u; \vect \Sigma)e^{-i\vect \omega^{T} \vect u } e^{ \frac{1}{2} \vect \omega^{T} \vect \Sigma \vect \omega}  e^{ -2\vect X^{T}\vect J_{n}^{T}\vect \Sigma \vect J_{n}\vect X } \nonumber \\
&\times & \frac{1}{(2\pi)^{n}\sqrt{\det(\vect \Sigma)}} \int_{\mathbb{R}^{2n}}d^{2n}\vect x \ e^{i\vect \omega^{T} \vect x}e^{ -\frac{1}{2}(\vect x-2i\vect \Sigma\vect J_{n} \vect X)^{T}\vect \Sigma^{-1} (\vect x-2i\vect \Sigma\vect J_{n}\vect X )}\label{AEA51}\\
&=&\frac{1}{(2\pi)^{n}}\int_{\mathbb{R}^{2n}}\int_{\mathbb{R}^{2n}}d^{2n}\vect \omega d^{2n}\vect u   \ p(\vect u; \vect \Sigma)e^{-i\vect \omega^{T} \vect u } e^{ \frac{1}{2} \vect \omega^{T} \vect \Sigma \vect \omega}   \nonumber \\
&\times & \frac{1}{(2\pi)^{n}\sqrt{\det(\vect \Sigma)}} \int_{\mathbb{R}^{2n}}d^{2n}\vect x \ e^{ -\frac{1}{2}\vect x^{T}\vect \Sigma^{-1} \vect x+i(\vect \omega+2\vect J_{n} \vect X)^{T} \vect x}\label{AEA52} \\
&=& \frac{1}{(2\pi)^{n}} \int_{\mathbb{R}^{2n}}\int_{\mathbb{R}^{2n}}d^{2n}\vect \omega d^{2n}\vect u \ p(\vect u; \vect \Sigma)e^{-i\vect \omega^{T} \vect u }e^{ \frac{1}{2} \vect \omega^{T} \vect \Sigma \vect \omega}  \left(e^{-\frac{1}{2}(\vect \omega +2 \vect J_{n} \vect X)^{T}\vect \Sigma (\vect \omega +2\vect J_{n} \vect X)}    \right) \label{AEA53} \\
&=& e^{ -2\vect X^{T}\vect J_{n}^{T}\vect \Sigma \vect J_{n}\vect X } \int_{\mathbb{R}^{2n}}d^{2n}\vect \omega \ e^{-2\vect \omega^{T} \vect \Sigma \vect J_{n} \vect X }   \left( \frac{1}{(2\pi)^{n}} \int_{\mathbb{R}^{2n}}d^{2n}\vect u \ e^{-i\vect \omega^{T} \vect u }p(\vect u; \vect \Sigma)\right) ,
\label{AEA5}
\end{eqnarray}
\end{widetext}
as we wanted to show. To derive Eq.(\ref{AEA5}), one can separate $\vect x$-dependent function from functions that depend on $\vect \omega$ and $\vect u$ only. Using $\vect \Sigma=\vect \Sigma^T$, one then arrives at Eq.(\ref{AEA52}). Performing the integration of $\vect x$ results in  
Eq.(\ref{AEA53}), rearranging terms yields to the desired form  Eq.(\ref{AEA5}). Substituting the explicit expression $p(\vect u; \vect \Sigma)=\frac{exp\left(-\frac{1}{2}\vect u^{T}(\vect V+\vect \Sigma)^{-1}\vect u \right) }{(2\pi)^{n}\sqrt{\det(\vect V+\vect \Sigma)}}$ in Eq. (\ref{AEA5}), and performing the integrals, we recover
\begin{equation}
\left|\Braket{\Phi_{1}| \hat{\rho}|\Phi_{2}  }\right|=\frac{e^{-2\vect X^{T}\vect J_{n}^{T} \frac{1}{\vect \Sigma^{-1} +\vect V^{-1}} \vect J_{n}\vect X }}{\sqrt{\det\left(\vect \Sigma +\vect V \right) }},
\nonumber 
\end{equation}
which is the first term in Eq.(\ref{Stau}).

%\bibliographystyle{apsrev}
%\bibliography{Library.bib}
%\bibliography{/home/avalido/Desktop/library/Library}
%\bibliography{C:/Users/dalonso/Documents/Research/Bibliography/References4_25022013}
%Merlin.mbs v4.21 2009-07-09.
%

\end{document}